\documentclass[11pt,a4paper]{article}
\usepackage[margin=2.35cm]{geometry}
\usepackage[T1]{fontenc}
\usepackage{lmodern}
\usepackage{microtype}
\usepackage{graphicx}
\usepackage{booktabs}
\usepackage{amsmath}
\usepackage{amssymb}
\usepackage{natbib}
\usepackage{caption}
\usepackage{subcaption}
\usepackage{xcolor}
\usepackage{authblk}
\usepackage{titlesec}
\usepackage[hidelinks]{hyperref}
\usepackage{needspace}
\usepackage{etoolbox}
\usepackage{enumitem}

\definecolor{paperink}{HTML}{30303C}
\definecolor{accent}{HTML}{4A4A5E}
\definecolor{rulegray}{HTML}{C9CAD2}
\hypersetup{colorlinks=true,linkcolor=accent,citecolor=accent,urlcolor=accent}
\color{paperink}
\titleformat{\section}{\large\bfseries\color{accent}}{\thesection}{0.65em}{}
\titleformat{\subsection}{\normalsize\bfseries\color{accent}}{\thesubsection}{0.65em}{}
\pretocmd{\section}{\needspace{6\baselineskip}}{}{}

\graphicspath{{figures/}}

\title{\bfseries Exploring Optimal Parameters for Ligand-Based Virtual Screening in Early Drug Discovery}
\author[1,2,*]{Temitope Sobodu}
\author[3]{Victor Chibuzor Johnson}
\author[4]{Ryan Kern}
\author[5]{Peter Oni}
\affil[1]{Attention Labs, Boston, Massachusetts, USA}
\affil[2]{Boston Children's Hospital and Harvard Medical School, Boston, Massachusetts, USA}
\affil[3]{Florida Institute of Technology, Melbourne, Florida, USA}
\affil[4]{Georgia Institute of Technology, Atlanta, Georgia, USA}
\affil[5]{Worcester Polytechnic Institute, Worcester, Massachusetts, USA}
\affil[*]{Correspondence: \href{mailto:temi@attentionlab.ai}{temi@attentionlab.ai} \textbar{} \url{https://attentionlab.ai}}
\date{August 2026}

\begin{document}
\maketitle

\renewcommand{\footnoterule}{}
\begingroup
\renewcommand{\thefootnote}{}
\footnotetext{%
  \fontsize{5}{6.2}\selectfont
  \textbf{Author affiliations:}\\
  Temitope Sobodu: Attention Labs, Boston, Massachusetts, USA; Boston Children's Hospital and Harvard Medical School, Boston, Massachusetts, USA.\\
  Victor Chibuzor Johnson: Department of Chemistry and Chemical Engineering, Florida Institute of Technology, Melbourne, Florida, USA.\\
  Ryan Kern: School of Chemistry and Biochemistry, Georgia Institute of Technology, Atlanta, Georgia, USA.\\
  Peter Oni: Department of Chemistry and Biochemistry, Worcester Polytechnic Institute, Worcester, Massachusetts, USA.%
}
\endgroup
\addtocounter{footnote}{-1}

\begin{abstract}
Ligand-based virtual screening depends on choices that are often treated as implementation details, including the similarity threshold, fingerprint setting and atom-invariant scheme. We examined how these choices altered the composition of ranked searches against the Enamine library for four aminergic reference ligands: atomoxetine, bupropion, mirtazapine and venlafaxine. Candidate sets were evaluated by compound-weighted scaffold novelty, normalized Shannon scaffold diversity and three computational estimates of synthetic accessibility. We first identified ligand-specific operating points along cumulative Tanimoto-ranked searches. We then compared five extended-connectivity fingerprint settings at matched retrieval depths and compared ECFP4 with the feature-class analogue FCSFP4. Finally, 240 records, corresponding to 229 unique structures, were scored independently by three chemists who were blinded to the computational scores. No single Tanimoto cutoff described all four searches. ECFP8 provided the most stable pooled setting, although venlafaxine favored ECFP2. ECFP4 was the stronger primary fingerprint in pooled comparisons, whereas FCSFP4 contributed nonredundant chemical space. Agreement among individual chemists was moderate, and the mean rating was more reliable than a single rating. SCScore showed the highest association with the blinded consensus, but performance varied by ligand. These results support a staged screening design in which a pooled default is followed by ligand-specific calibration and expert review.
\end{abstract}

\section{Introduction}
The number of synthetically plausible, drug-like molecules exceeds the number that can be prepared and tested experimentally by many orders of magnitude \citep{polishchuk2013}. Virtual screening reduces this search space by ranking compounds before experimental evaluation \citep{shoichet2004,stumpfe2012}. Structure-based methods use an experimentally determined or modeled target structure, whereas ligand-based virtual screening derives its ranking from one or more reference ligands and does not require an explicit receptor model \citep{stumpfe2012,vonkorff2009}. Ligand-based methods are therefore useful when a relevant ligand is known but structural information for the target is unavailable or insufficient for a large docking campaign \citep{stumpfe2012}. They can also complement structure-based methods because the two strategies encode different information and can prioritize partly distinct chemotypes \citep{vonkorff2009}. In both settings, virtual screening is an enrichment procedure intended to concentrate useful candidates near the top of a ranked list, rather than a claim that every library member has been placed in its true biological order \citep{shoichet2004,truchon2007}.

\begin{figure}[!htbp]
\centering
\includegraphics[width=0.92\textwidth]{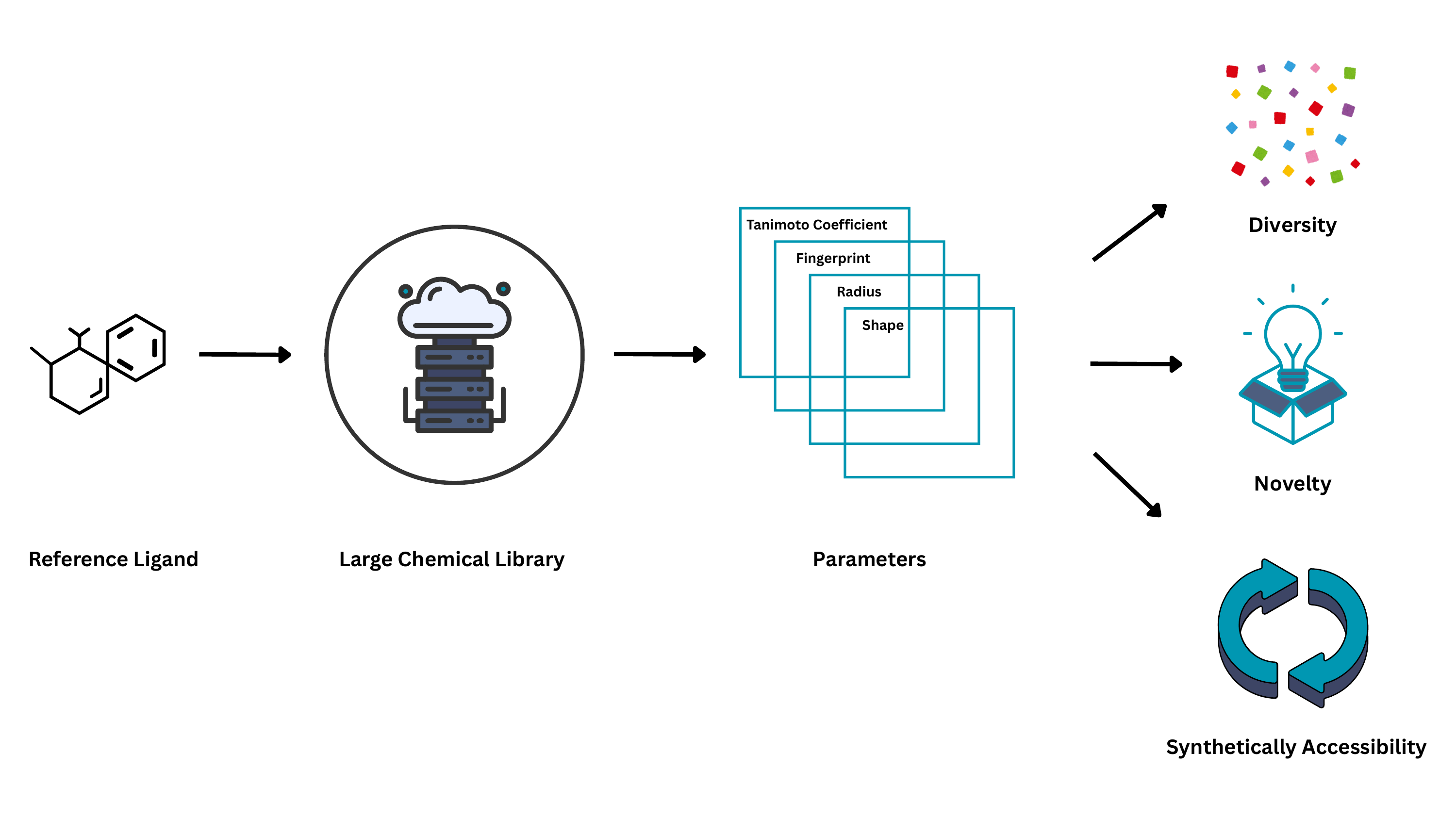}
\captionsetup{width=0.92\textwidth}
\caption{Graphical abstract. Parameter-selection workflow for ligand-based virtual screening. Four reference ligands were used to query the Enamine library while varying the Tanimoto threshold, extended-connectivity fingerprint setting and atom-invariant scheme. Retrieved sets were evaluated for scaffold novelty, scaffold diversity and predicted synthetic accessibility, followed by comparison of computational accessibility scores with blinded chemist judgments.}
\label{fig:graphical-abstract}
\end{figure}

Similarity searching is a direct form of ligand-based virtual screening in which compounds are ranked by their resemblance to a query molecule \citep{willett2006}. Its rationale is the similar-property principle, under which structurally related molecules are more likely than randomly selected molecules to share physicochemical or biological properties \citep{martin2002,willett2006}. The relationship is probabilistic rather than deterministic. Small structural changes can alter potency or other biological properties, and activity cliffs are well-established counterexamples to a simple mapping between structural and biological similarity \citep{martin2002}. A numerical similarity value consequently reports agreement under a defined representation and comparison rule, not biological equivalence \citep{riniker2013,baldi2010}. The operational output is a ranked chemical neighborhood around the query whose membership depends on the descriptor, similarity measure, reference ligand and screening collection \citep{hert2006,riniker2013}.

Two-dimensional fingerprints are widely used in this setting because they provide compact molecular representations that can be compared rapidly across large libraries \citep{willett2006,riniker2013}. A binary fingerprint maps structural features to bits, with each set bit denoting the presence of an encoded molecular environment \citep{willett2006}. Extended-connectivity fingerprints, or ECFPs, enumerate circular atom environments by iteratively expanding from each atom through the molecular graph \citep{rogers2010}. In the conventional ECFP nomenclature, the numerical suffix denotes the diameter of the largest encoded environment, so ECFP4 ordinarily corresponds to environments generated to a radius of two bonds \citep{rogers2010}. Smaller settings emphasize local atom neighborhoods, whereas larger settings encode more surrounding topology and can separate compounds that share only short fragments \citep{rogers2010,sastry2010}. Fingerprint family and parameter choices can materially change enrichment and ranking behavior, even when the query and database are held constant \citep{sastry2010,riniker2013}. Because software conventions are not completely uniform, the present analysis preserves the supplied labels ECFP0, ECFP2, ECFP4, ECFP6 and ECFP8 and treats them as ordered fingerprint settings.

The atom invariants used to initialize a circular fingerprint provide a second level of control. Standard ECFPs distinguish atoms using properties that include element, valence, charge, hydrogen count and aromaticity \citep{rogers2010}. Functional-class fingerprints replace part of this atom-specific description with generalized chemical roles such as hydrogen-bond donor, hydrogen-bond acceptor, aromatic, hydrophobic, positively ionizable and negatively ionizable character \citep{rogers2010}. In this study, FCSFP4 denotes the functional-class analogue of ECFP4. Functional-class encoding can assign related features to chemically distinct atoms that occupy similar interaction roles, thereby supporting retrieval beyond exact atom-level substructures \citep{rogers2010}. This behavior is relevant to scaffold hopping, for which two-dimensional fingerprints can retrieve alternative frameworks while retaining selected local features \citep{vogt2010}. It may also broaden the neighborhood and reduce atom-specific correspondence, so whether FCSFP4 should replace or complement ECFP4 must be determined empirically for the query series.

For binary fingerprints, molecular similarity is commonly summarized by the Tanimoto coefficient \citep{willett2006,bajusz2015}. If $A$ and $B$ are the sets of on-bits for two molecules, their similarity is
\begin{equation}
T(A,B)=\frac{|A\cap B|}{|A\cup B|}.
\end{equation}
The coefficient ranges from 0 for fingerprints with no shared on-bits to 1 for identical bit sets, and its statistical behavior supports its use with sparse binary fingerprints \citep{bajusz2015}. A restrictive cutoff concentrates a search on close analogues but can return a small or scaffold-redundant set, whereas a permissive cutoff expands coverage while increasing the chance that shared fingerprint features are insufficient to preserve relevant biology \citep{willett2006,vogt2010}. The distribution and interpretation of Tanimoto values depend on fingerprint composition and the chemical population being compared \citep{baldi2010}. The same numerical cutoff therefore need not define an equivalent neighborhood across ECFP settings or between ECFP and FCSFP representations \citep{sastry2010,riniker2013}. A reported threshold is most informative when accompanied by the fingerprint definition, query ligand and number of retained compounds.

Parameter selection in similarity searching consequently balances local exploitation against broader chemical exploration \citep{willett2006,vogt2010}. Retrieval count alone does not resolve this balance because a large set may contain many close analogues of one chemotype, while a smaller set may span more distinct frameworks \citep{krier2006}. Maximizing structural difference without considering precedent or synthetic feasibility can also produce candidates that are difficult to progress \citep{skoraczynski2023}. We therefore evaluated three properties of each retrieved set: novelty, diversity and synthetic accessibility.

Novelty is used here in a defined cheminformatics sense. Each compound was reduced to a scaffold representation based on the molecular-framework concept of ring systems and their connecting linkers \citep{bemis1996}. A scaffold was designated novel when it was absent from the supplied WIPO-derived patent annotation, and compound-weighted novelty was calculated as the fraction of evaluated compounds assigned to such scaffolds. The metric records structural nonoccurrence relative to the indexed collection and remains conditional on database coverage, structure standardization and scaffold definition \citep{bemis1996,krier2006}. It is not a legal determination of patent novelty or freedom to operate. Within these limits, it provides a reproducible measure of whether a search remains in previously represented framework space or extends beyond it.

Diversity describes how broadly a set is distributed across a specified representation of chemical space \citep{krier2006}. The primary unit in this study was the scaffold rather than the complete molecular graph. Unique scaffold count records framework richness but cannot distinguish a balanced collection from one dominated by a single scaffold series \citep{krier2006}. We therefore used normalized Shannon entropy, which incorporates both the number of scaffold classes and the evenness of their frequencies \citep{shannon1948,krier2006}. A higher value indicates broader and more even scaffold representation. Compound and scaffold overlap were analyzed separately to establish whether fingerprint searches sampled the same structures or nonredundant chemical neighborhoods.

Synthetic accessibility refers to the expected difficulty of preparing a molecule from available starting materials through plausible transformations \citep{ertl2009,skoraczynski2023}. It is distinct from structural novelty, biological activity and general developability, and it is not directly observed until synthesis is attempted \citep{skoraczynski2023}. Computational scores provide rapid proxies but encode different targets and are not interchangeable \citep{skoraczynski2023}. The SA score combines fragment contributions derived from known compounds with structural-complexity penalties and ordinarily increases with predicted difficulty \citep{ertl2009}. RAscore is a machine-learned surrogate for whether the AiZynthFinder retrosynthetic planner can identify a route, with larger values indicating greater predicted accessibility \citep{thakkar2021}. SCScore learns synthetic complexity from reaction data by modeling the tendency for products to be more complex than reactants, with larger values indicating greater predicted complexity \citep{coley2018}. SA score therefore emphasizes fragment precedent and molecular complexity, RAscore approximates a retrosynthetic search outcome, and SCScore represents reaction-derived complexity \citep{ertl2009,thakkar2021,coley2018}. Their agreement is informative, but disagreement is expected because the scores were trained or constructed against different operational definitions of synthesizability \citep{skoraczynski2023}.

These criteria define a multiobjective selection problem. Tightening a similarity threshold, expanding the encoded ECFP environment or replacing atom invariants with functional classes can improve one outcome while reducing another. A setting may increase scaffold novelty at the expense of retrieval size, or improve one accessibility proxy without improving the others. An optimum is therefore conditional on the stated objectives and their relative weights. We evaluated Pareto efficiency, effect sizes, rank stability across retrieval depths and sensitivity to alternative weights assigned to novelty, diversity and synthesis. These quantities formed the primary decision framework, while nominal tests on correlated cumulative sets were treated as descriptive.

Here we examined ranked searches of the Enamine library initiated from four aminergic drugs: atomoxetine, bupropion, mirtazapine and venlafaxine. These ligands differ in topology and functional-group arrangement, providing four query contexts within a defined pharmacological domain. The analysis was restricted to the composition and prioritization of the retrieved chemical space. Biological activity, target engagement and prospective synthesis outcome were not measured. Four questions were specified. First, when ECFP4 is fixed, does a common Tanimoto cutoff provide a stable balance among compound-weighted scaffold novelty, scaffold diversity and predicted synthetic accessibility, or are ligand-specific operating points required? Second, when retrieval depth is matched, which ECFP setting provides the most stable performance across the four ligands and across alternative scientific priorities? Third, does FCSFP4 provide a sufficient improvement to replace ECFP4, or does it function more appropriately as a complementary search that adds nonredundant compounds and scaffolds? Fourth, how closely do SA score, RAscore and SCScore correspond to an independent consensus from three chemists who assigned synthetic-difficulty ratings without seeing the computational scores? Together, these questions test whether a transferable parameter set can be identified and where ligand-specific calibration remains necessary.

\begin{figure}[!htbp]
\centering
\includegraphics[width=0.88\textwidth]{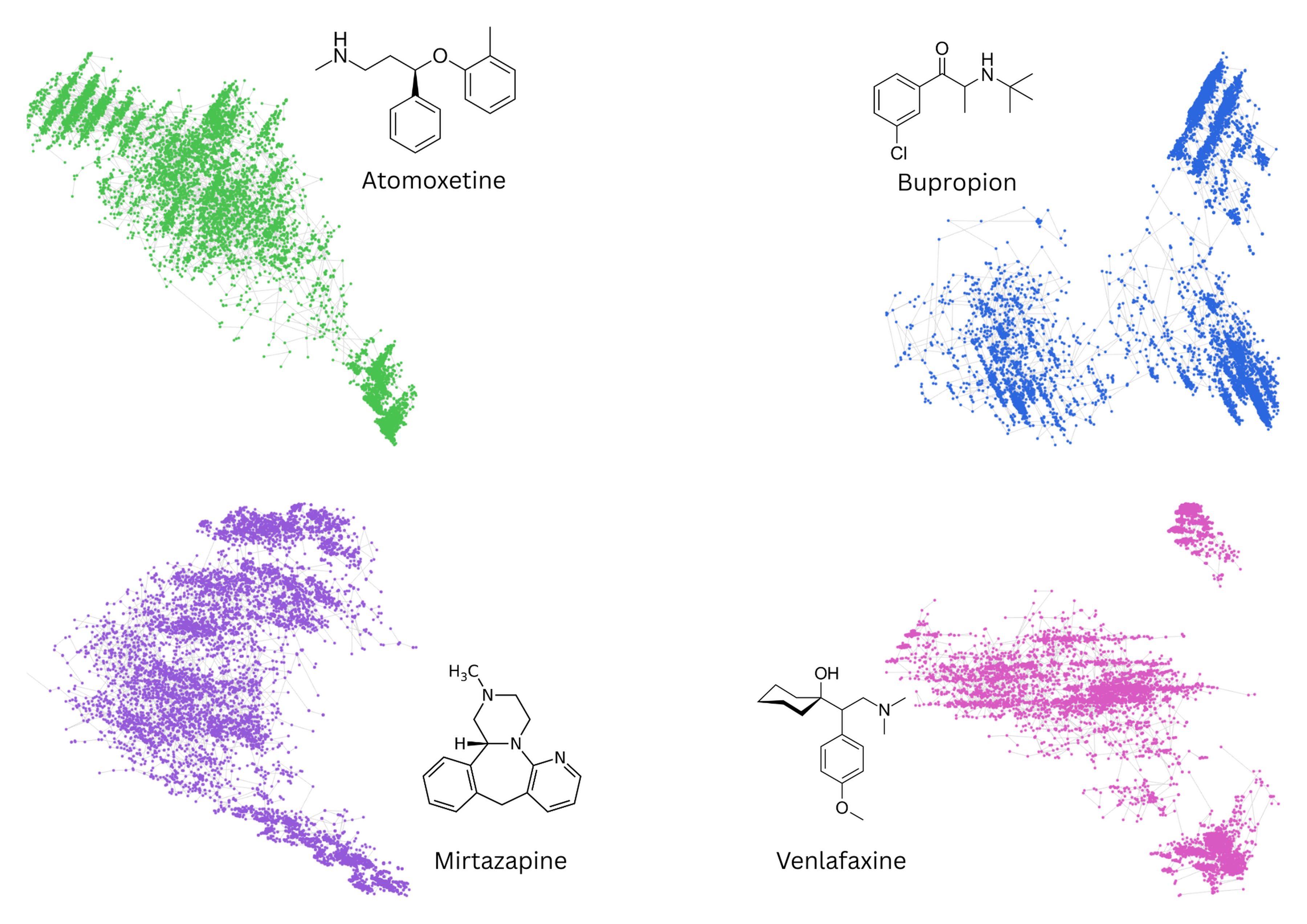}
\captionsetup{width=0.92\textwidth}
\caption{Ligand-centered organization of the retrieved chemical space. Two-dimensional TMAP projections are shown for the screening outputs initiated from atomoxetine, bupropion, mirtazapine and venlafaxine, together with the corresponding query structures. Points denote retrieved compounds, connecting lines follow the neighborhood graph used in the projection. An \href{https://bbbnukeestorage7791b0f55.z13.web.core.windows.net/adhd-synth-tmap/index.html}{interactive TMAP} is available online.}
\label{fig:ligand-tmap}
\end{figure}

\section{Results}
\subsection{Tanimoto operating points were ligand dependent}
For each ECFP4 search, cumulative sets were evaluated at 100-compound increments. A threshold retained all compounds with similarity at or above the minimum value in that cumulative set. Sets containing fewer than 1,000 compounds were excluded from selection. The eligible points were evaluated in a three-objective space comprising compound-weighted WIPO scaffold novelty, normalized Shannon scaffold diversity and a synthesis composite. The selected operating point maximized equal-weight standardized utility among Pareto-efficient candidates.

Pareto screening first removed thresholds that were no better on any of the three objectives. Equal-weight standardized utility then selected among the remaining nondominated trade-offs, as defined in Methods. This sequence separated the exclusion of objectively inferior thresholds from the preference applied among viable alternatives.

The selected minimum Tanimoto coefficients were 0.265 for atomoxetine, 0.415 for bupropion, 0.224 for mirtazapine and 0.324 for venlafaxine (Table~\ref{tab:tc}). Atomoxetine, mirtazapine and venlafaxine retained 10,000 compounds at their selected points. Bupropion retained 3,800 compounds. The near-optimal intervals were narrow for atomoxetine and venlafaxine, although both values occurred at the broadest available retrieval boundary. The mirtazapine interval extended from 0.224 to 0.317, indicating that the available outcomes did not identify a precise cutoff. Bupropion showed an interior operating point at 0.415.

\begin{table}[htbp]
\centering
\caption{Ligand-specific ECFP4 operating points. Near-optimal intervals contain Pareto-efficient candidates within 0.15 standardized utility units of the maximum.}
\label{tab:tc}
\small
\begin{tabular}{lrrrrl}
\toprule
Reference ligand & Minimum TC & Retained & Novelty (\%) & $D$ & Near-optimal interval \\
\midrule
Atomoxetine & 0.265 & 10,000 & 37.42 & 0.528 & 0.265--0.269 \\
Bupropion & 0.415 & 3,800 & 34.00 & 0.437 & 0.407--0.415 \\
Mirtazapine & 0.224 & 10,000 & 95.45 & 0.646 & 0.224--0.317 \\
Venlafaxine & 0.324 & 10,000 & 68.87 & 0.639 & 0.324--0.329 \\
\bottomrule
\end{tabular}
\end{table}

\begin{figure}[!t]
\centering
\includegraphics[width=0.92\textwidth]{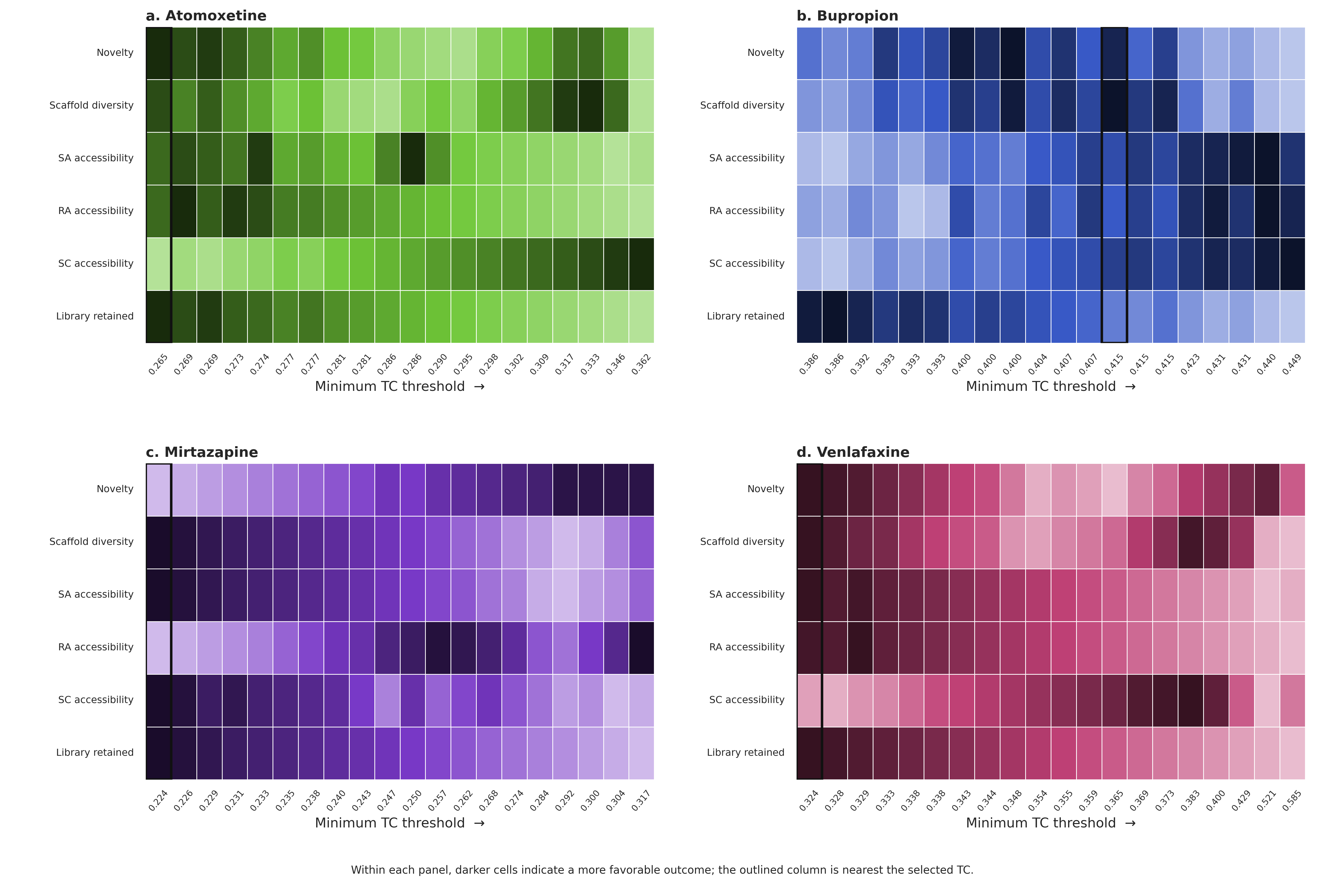}
\caption{Outcome profiles across cumulative ECFP4 similarity thresholds. Each panel reports within-ligand changes in compound-weighted novelty, normalized Shannon scaffold diversity and the synthesis composite. The outlined region denotes the selected operating point. Colors are scaled within each ligand and should not be compared as absolute values between panels.}
\label{fig:tc}
\end{figure}

The response surfaces showed why a common absolute threshold was not supported (Fig.~\ref{fig:tc}). Tightening the cutoff reduced coverage in every search, but the corresponding changes in novelty, diversity and synthesis were not parallel across ligands. In the bupropion search, the intermediate threshold retained 38\% of the ranked output and yielded 34.0\% compound-weighted novelty with 589 unique scaffolds. Mirtazapine remained highly novel over a broad interval, so a narrower numerical optimum would have overstated the resolution of the data.

Smoothed trajectories were used to summarize curve shape without extrapolation beyond the observed thresholds. Pseudo-$R^2$ values ranged from 0.920 to 0.997 across the six outcomes and four ligands. Spearman correlations between threshold and compound-weighted novelty were $-0.463$ for atomoxetine, $-0.441$ for bupropion, 0.992 for mirtazapine and $-0.335$ for venlafaxine. The corresponding utility correlations were $-0.334$, 0.355, $-0.884$ and $-0.662$. Thus, neither the direction nor the magnitude of the novelty-utility trade-off was conserved across ligand searches. Pareto-efficient solutions within 0.15 standardized utility units of the maximum retained 9,500--10,000 compounds for atomoxetine, 3,700--5,100 for bupropion, 1,000--10,000 for mirtazapine and 9,100--10,000 for venlafaxine. These ranges quantify the resolution with which a threshold could be identified from the observed objectives.

\subsection{ECFP8 was the most stable pooled setting, with a ligand interaction}
ECFP0, ECFP2, ECFP4, ECFP6 and ECFP8 were compared at matched top-$N$ depths of 1,000, 2,500, 5,000, 7,500 and 10,000 compounds. Across ligands and depths, ECFP8 had the lowest mean rank, 2.15, and was Pareto-efficient in 95\% of comparisons. It was selected in 74.1\% of 5,000 randomly weighted novelty-diversity-synthesis priority schemes. ECFP2 had a mean rank of 2.80, a Pareto rate of 80\% and an overall winner share of 25.4\%.

Robustness was summarized by mean utility rank and the proportion of the 20 ligand-depth strata in which a setting was Pareto-efficient. ECFP8 had a mean utility of 0.276 standardized units, compared with 0.193 for ECFP2, 0.094 for ECFP4, 0.073 for ECFP6 and $-0.636$ for ECFP0. ECFP8 was the top equal-weight setting in five strata and had a worst observed rank of five, indicating that its pooled advantage arose from favorable average performance rather than universal first-place ranking.

The pooled ordering did not describe every ligand. ECFP8 was selected in 60.8\% of weighting schemes for atomoxetine and 56.8\% for bupropion. For mirtazapine, ECFP0 and ECFP8 accounted for 52.7\% and 46.4\% of selections, respectively. ECFP2 ranked first at every matched depth for venlafaxine and was selected in 88.8\% of weighting schemes (Fig.~\ref{fig:radius}). At the top 10,000 depth, ECFP2 retrieved 2,314 unique venlafaxine scaffolds and 77.74\% compound-weighted novelty.

A linear response model included fingerprint setting, ligand, their interaction and log-transformed retrieval depth. The interaction tested whether setting effects changed with the reference ligand after accounting for depth. For composite utility, the setting-by-ligand term had $F_{12,79}=18.05$, descriptive $P<0.001$ and partial $\eta_p^2=0.733$ (Table~\ref{tab:q2models}). The interaction accounted for a large fraction of the model-attributable variation in Shannon diversity and RAscore, and a smaller but still substantial fraction for novelty and SA score. SCScore alone did not show a clearly resolved interaction.

\begin{table}[htbp]
\centering
\caption{Setting-by-ligand terms from the matched-depth response models. The tests describe nested top-$N$ response surfaces and are not treated as independent-sample inference.}
\label{tab:q2models}
\small
\begin{tabular}{lrrr}
\toprule
Outcome & $F_{12,79}$ & Descriptive $P$ & Partial $\eta_p^2$ \\
\midrule
Compound-weighted novelty & 8.56 & $<0.001$ & 0.565 \\
Normalized Shannon diversity & 29.70 & $<0.001$ & 0.819 \\
SA score & 6.96 & $<0.001$ & 0.514 \\
RAscore & 20.83 & $<0.001$ & 0.760 \\
SCScore & 1.50 & 0.141 & 0.186 \\
Composite utility & 18.05 & $<0.001$ & 0.733 \\
\bottomrule
\end{tabular}
\end{table}

Weighting sensitivity was evaluated across 5,000 randomly sampled priority schemes spanning novelty, diversity and synthesis. The largest Monte Carlo standard error for a winner share was 0.71 percentage points. The observed 74.1\% share for ECFP8 and 25.4\% share for ECFP2 therefore exceeded the uncertainty attributable to the finite simulation. Because the ranked top-$N$ sets are cumulative, the regression terms and their $P$ values were retained as descriptive summaries.

\begin{figure}[!t]
\centering
\begin{subfigure}{0.88\textwidth}
\centering\includegraphics[width=\textwidth]{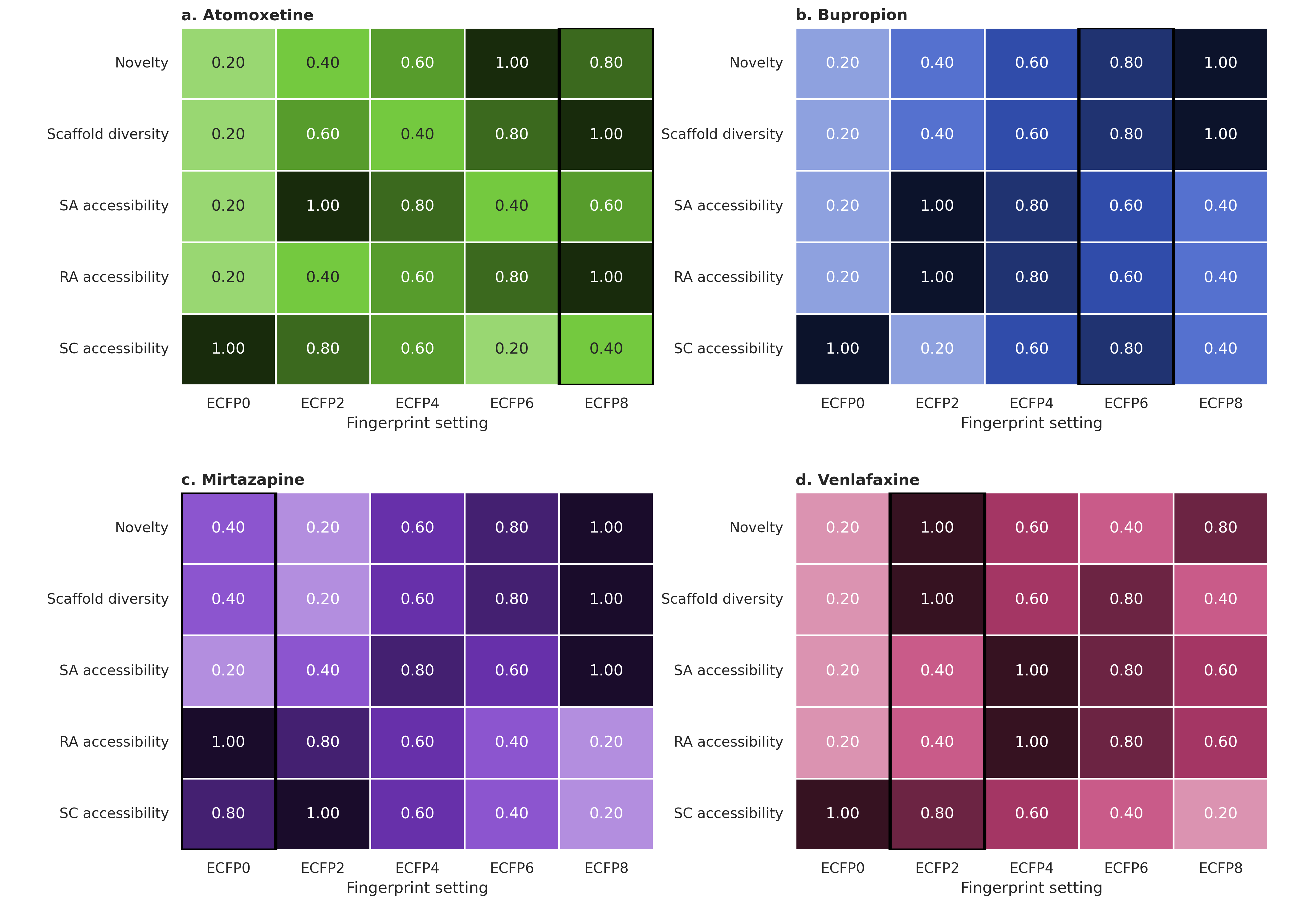}
\caption{Top-10,000 outcome profiles. Values are favorable-direction within-ligand percentiles. Outlines denote the most robust mean-rank setting.}
\end{subfigure}
\caption{Performance of ECFP settings. The outcome profiles show that the preferred setting depends on the ligand and evaluation criterion.}
\label{fig:radius}
\end{figure}

\begin{figure}[!t]\ContinuedFloat
\centering
\begin{subfigure}{0.78\textwidth}
\centering\includegraphics[width=\textwidth]{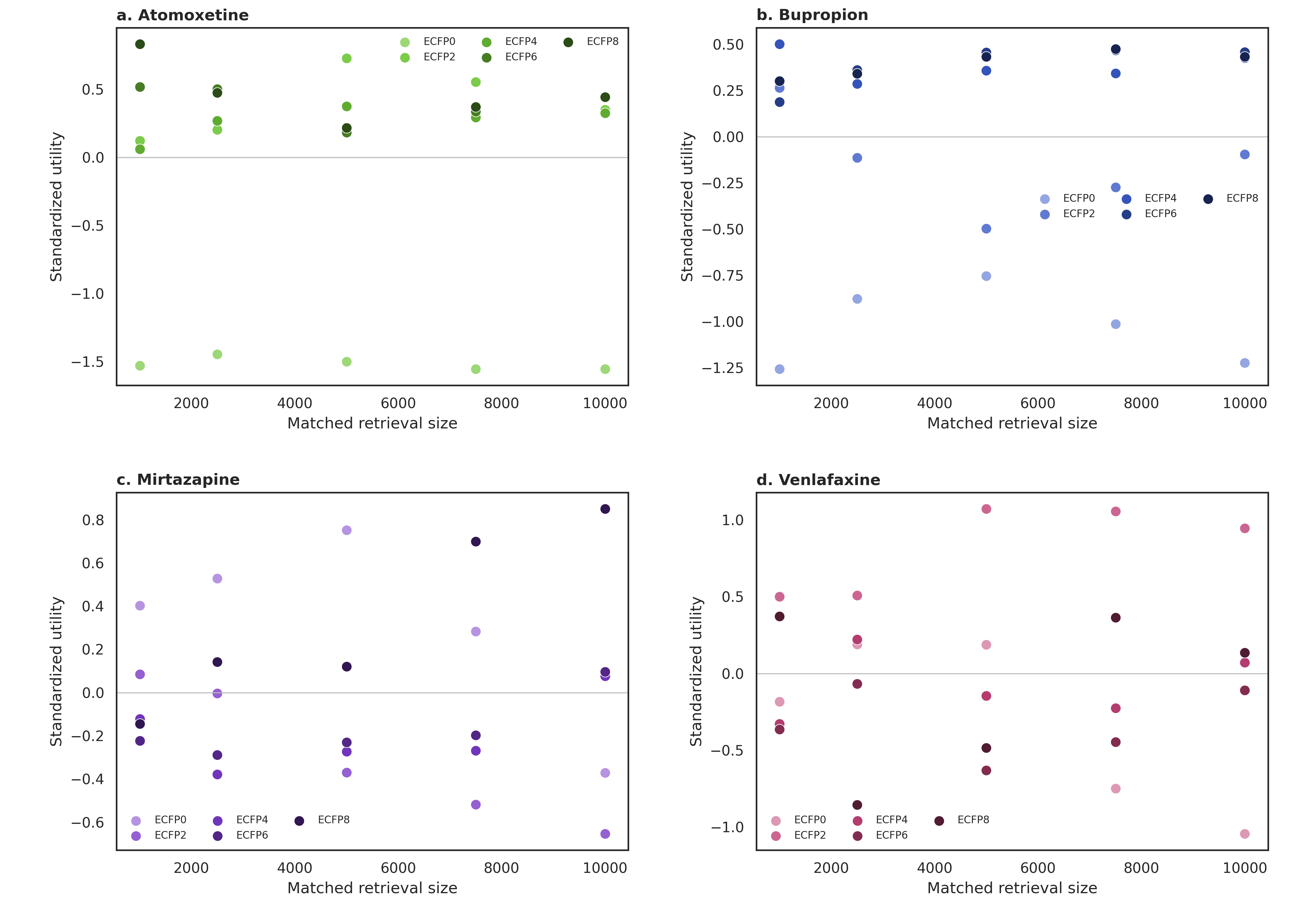}
\caption{Composite utility across matched retrieval depths.}
\end{subfigure}
\vspace{0.35em}
\begin{subfigure}{0.88\textwidth}
\centering\includegraphics[width=\textwidth]{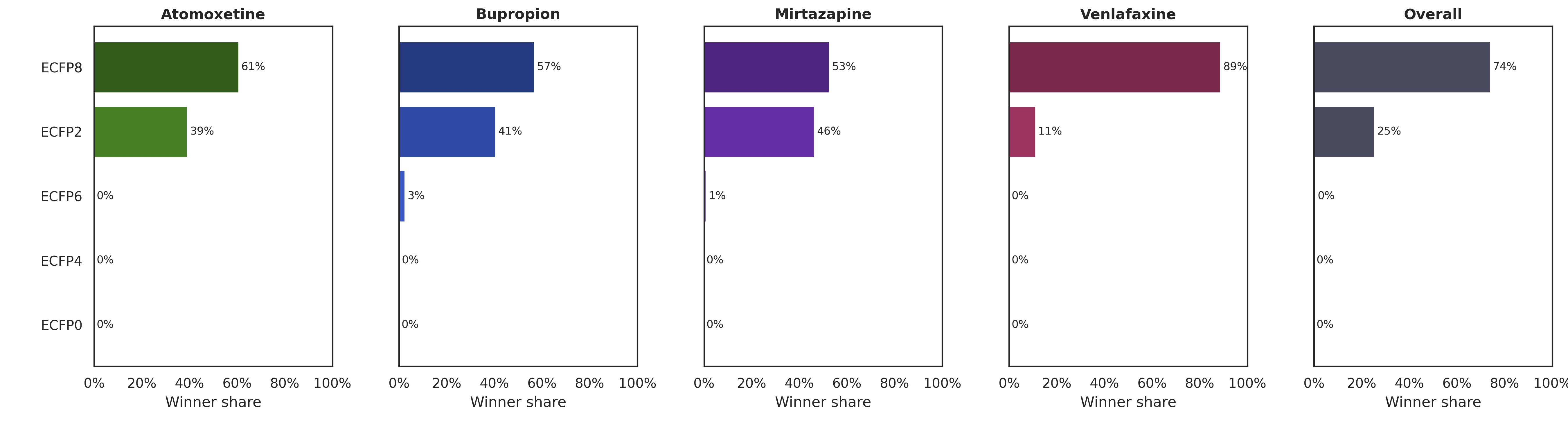}
\caption{Winner share across 5,000 randomly weighted priority schemes.}
\end{subfigure}
\caption[]{Performance of ECFP settings, continued. ECFP8 was the most stable pooled setting, whereas the venlafaxine search consistently favored ECFP2. The overall panel is shown in neutral gray.}
\end{figure}

\subsection{FCSFP4 sampled complementary space but did not replace ECFP4}
ECFP4 and FCSFP4 were compared at the same five ranked retrieval depths. Raw Tanimoto coefficients were not compared because their scales depend on the fingerprint representation. ECFP4 had higher equal-weight utility in 14 of 20 ligand-depth comparisons and was selected in 91.5\% of pooled weighting schemes. Its selection frequency was 62.6\% for atomoxetine, 86.1\% for bupropion and 95.5\% for mirtazapine. FCSFP4 was selected in 61.3\% of venlafaxine schemes.

Each FCSFP4 outcome was subtracted from its ECFP4 counterpart after matching on ligand and retrieval depth. Positive differences favor FCSFP4 for novelty, diversity and RAscore, whereas negative differences favor FCSFP4 for SA score and SCScore. Utility was already coded in a common favorable direction. The Wilcoxon signed-rank statistic was used as a descriptive summary of the 20 paired differences, with direction counts and mean differences retained as the primary measures because the five cumulative depths within each ligand are dependent.

Across the 20 matched comparisons, ECFP4 had higher compound-weighted novelty in 17. The mean FCSFP4-minus-ECFP4 difference was $-6.93$ percentage points, with a median difference of $-6.02$ percentage points and descriptive signed-rank $P=0.0007$ (Table~\ref{tab:q3paired}). ECFP4 also had higher utility in 14 comparisons; the mean standardized difference was $-0.533$ and the descriptive $P$ value was 0.0126. FCSFP4 had a higher RAscore in 13 comparisons, indicating that its disadvantage in pooled utility was not uniform across synthesis-related outcomes. The differences in Shannon diversity, SA score, RAscore and SCScore were less directionally consistent than the novelty difference.

\begin{table}[htbp]
\centering
\caption{Matched FCSFP4-minus-ECFP4 contrasts across 20 ligand-depth pairs. Counts indicate the fingerprint with the larger raw value; lower values are favorable for SA score and SCScore.}
\label{tab:q3paired}
\small
\begin{tabular}{lrrrr}
\toprule
Outcome & Mean $\Delta$ & Median $\Delta$ & FCSFP4 larger & Descriptive $P$ \\
\midrule
Compound-weighted novelty & $-0.0693$ & $-0.0602$ & 3/20 & 0.0007 \\
Normalized Shannon diversity & $-0.0112$ & $-0.0220$ & 7/20 & 0.3683 \\
SA score & 0.0362 & $-0.0412$ & 9/20 & 0.9563 \\
RAscore & 0.0112 & 0.0016 & 13/20 & 0.1231 \\
SCScore & $-0.0703$ & $-0.0304$ & 8/20 & 0.0583 \\
Composite utility & $-0.5333$ & $-0.8889$ & 6/20 & 0.0126 \\
\bottomrule
\end{tabular}
\end{table}

\begin{figure}[!t]
\centering
\begin{subfigure}{0.84\textwidth}
\centering\includegraphics[width=\textwidth]{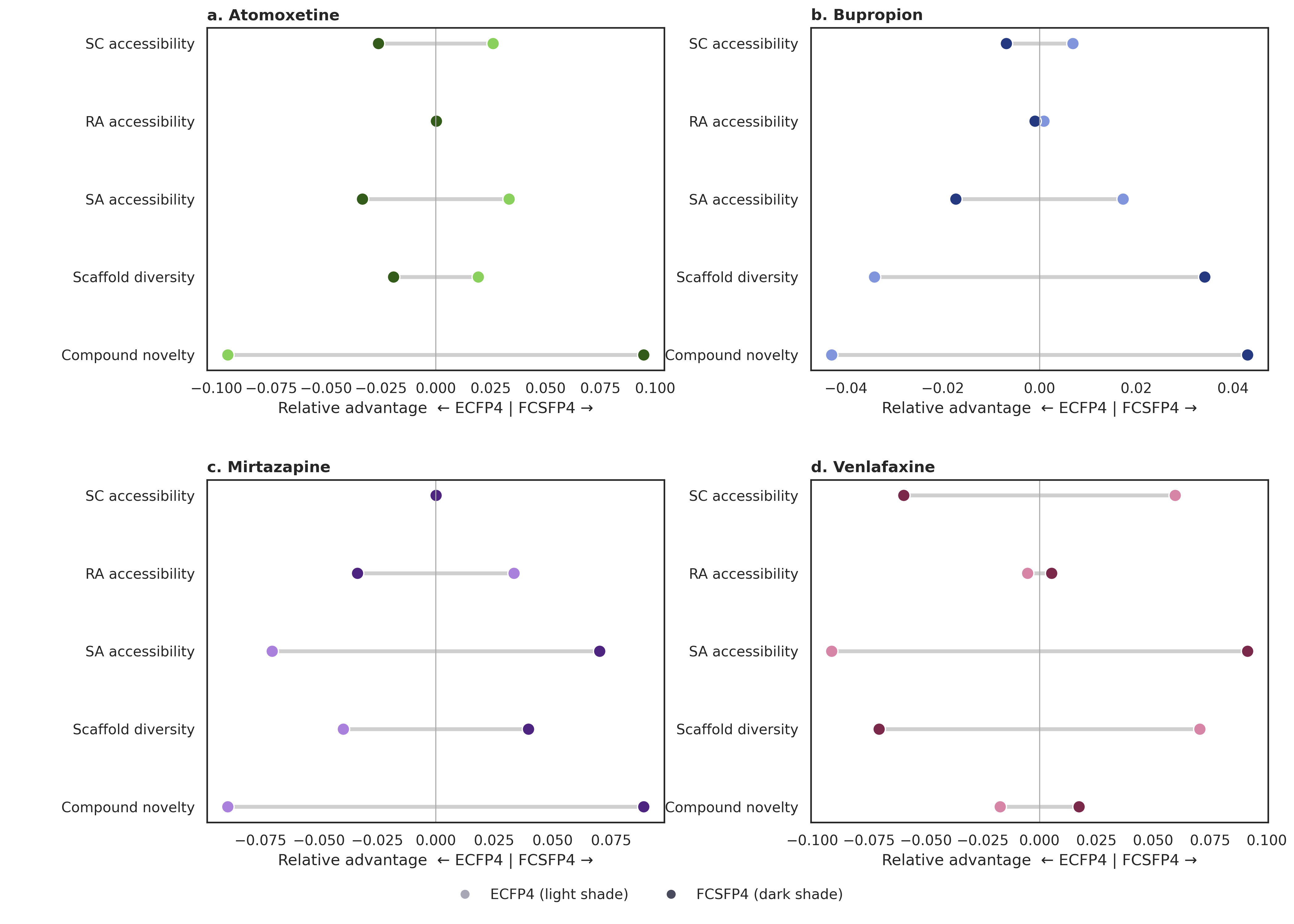}
\caption{Matched outcome differences.}
\end{subfigure}
\vspace{0.45em}
\begin{subfigure}{0.84\textwidth}
\centering\includegraphics[width=\textwidth]{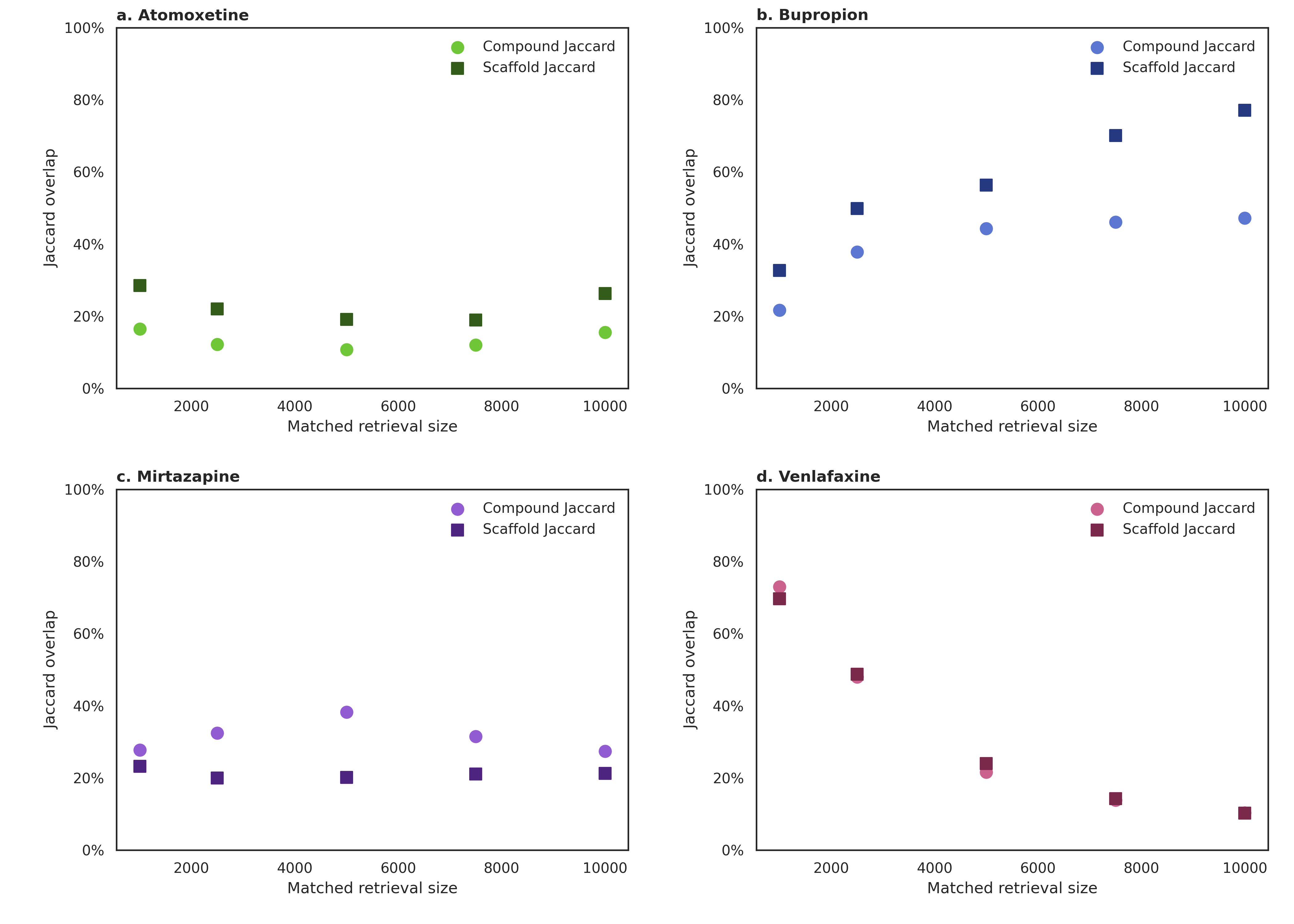}
\caption{Compound and scaffold overlap.}
\end{subfigure}
\caption{Comparison of ECFP4 and FCSFP4 at matched retrieval depths. ECFP4 more often preserved the pooled balance of novelty, diversity and synthesis, while FCSFP4 retrieved a nonredundant neighborhood whose magnitude depended on the reference ligand.}
\label{fig:fingerprint}
\end{figure}

The two fingerprints frequently retrieved different structures (Fig.~\ref{fig:fingerprint}). Compound and scaffold overlap were measured with the Jaccard index, defined as the size of the intersection divided by the size of the union. At the top 10,000 depth, compound overlap ranged from 10.3\% for venlafaxine to 47.2\% for bupropion. Scaffold overlap ranged from 10.3\% to 77.1\%. Adding nonredundant FCSFP4 results to the ECFP4 set contributed 7,310 compounds and 704 scaffolds for atomoxetine, 3,588 compounds and 115 scaffolds for bupropion, 5,697 compounds and 1,341 scaffolds for mirtazapine, and 8,134 compounds and 2,907 scaffolds for venlafaxine.

\begin{figure}[!t]
\centering
\begin{subfigure}{0.92\textwidth}
\centering\includegraphics[width=\textwidth]{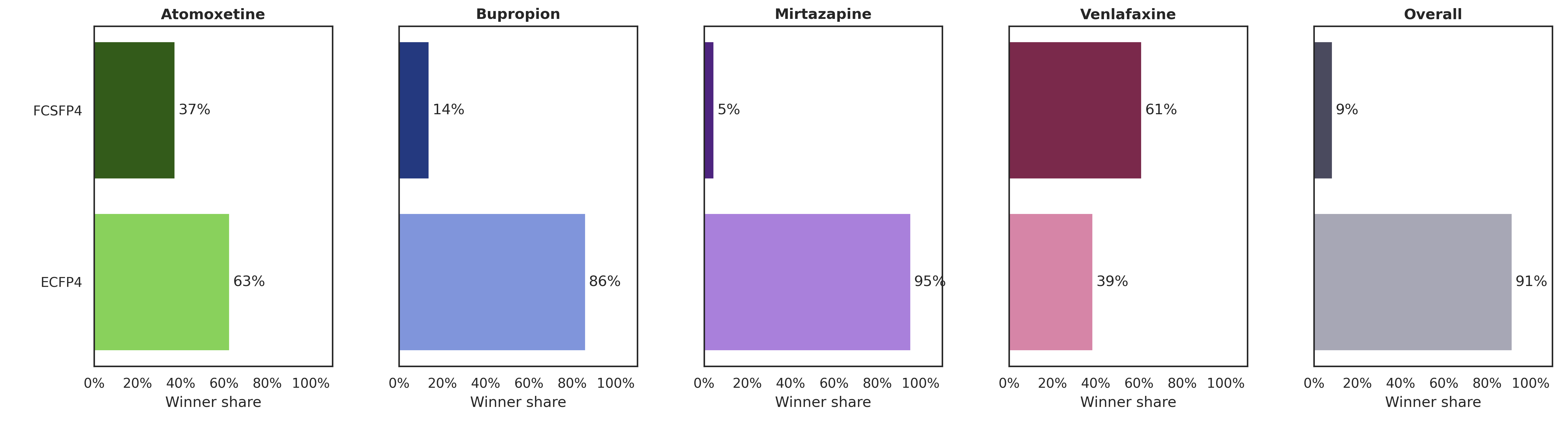}
\caption{Winner shares for ECFP4 and FCSFP4 under 5,000 alternative novelty-diversity-synthesis priority schemes.}
\end{subfigure}
\vspace{0.45em}
\begin{subfigure}{0.82\textwidth}
\centering\includegraphics[width=\textwidth]{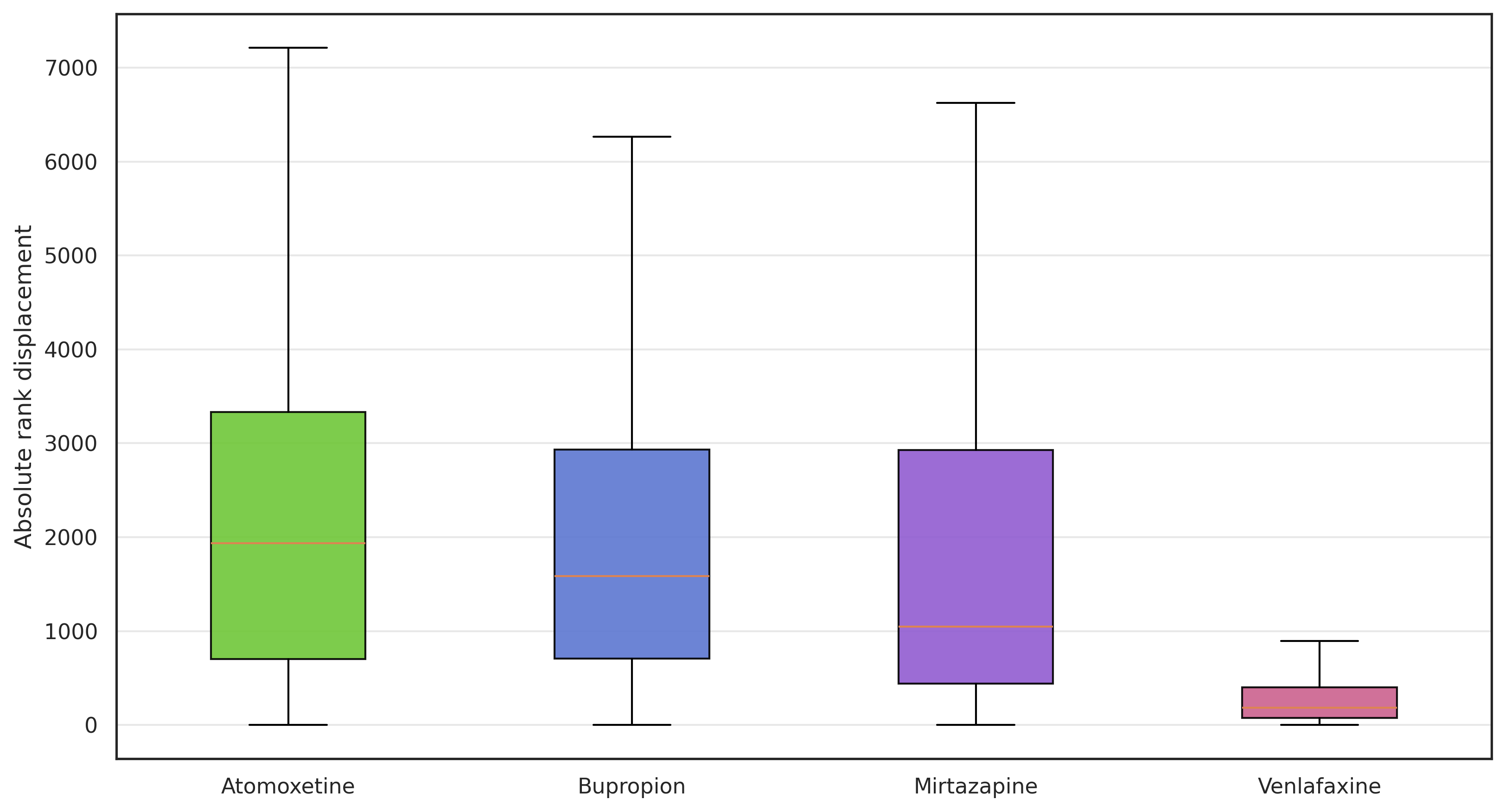}
\caption{Absolute rank displacement among compounds retrieved by both fingerprints.}
\end{subfigure}
\caption{Sensitivity and rank reordering in the ECFP4-FCSFP4 comparison. Pooled preference favored ECFP4, while venlafaxine favored FCSFP4. Shared structures frequently changed position even when their compound-level properties were unchanged.}
\label{fig:fingerprint_sensitivity}
\end{figure}

Reordering of a shared compound was measured as the absolute difference between its ECFP4 and FCSFP4 ranks. Median displacements were 1,937 ranks for atomoxetine, 1,585 for bupropion, 1,049 for mirtazapine and 186 for venlafaxine. Spearman correlations between the two rank lists among shared compounds were 0.812, 0.598, 0.362 and 0.910, respectively. Thus, low set overlap and large rank displacement were distinct properties: venlafaxine had the lowest compound overlap but the most concordant ordering among the relatively small subset retrieved by both representations. For shared structures, the synthesis scores were identical; only retrieval ranks differed. Weight sensitivity and rank displacement are shown in Fig.~\ref{fig:fingerprint_sensitivity}.

\subsection{Blinded chemist judgments provided an external synthesis benchmark}
Three chemists independently assigned synthetic-difficulty scores without access to SA score, RAscore or SCScore. The merged audit set contained 240 records across ECFP0, ECFP2, ECFP4, ECFP6, ECFP8 and FCSFP4. These records represented 229 unique molecular structures. Six structures occurred more than once, accounting for 17 records. The 229-structure set was used for score validation, while the 240-record set was retained for agreement and repeatability analyses.

The chemists used different numerical ranges. Ratings were therefore converted to within-rater percentile ranks and standardized values before aggregation. Higher values denoted greater perceived difficulty. Pairwise Spearman correlations were 0.366, 0.353 and 0.314. Kendall's coefficient of concordance was 0.556. The absolute-agreement intraclass correlation was 0.358 for a single rater and 0.625 for the mean of three raters. Among repeated structures, the mean within-structure standard deviations on the normalized raw scales were 0.050, 0.091 and 0.069 for the three chemists.

Within-rater ranks were converted to percentiles using average ranks for ties and then averaged across the three chemists to form the primary ordinal consensus. Standardized ratings were averaged separately for calibration analyses. Kendall's $W$ was calculated from summed within-rater ranks, and absolute agreement was estimated with a two-way random-rater ICC. The increase from ICC$(2,1)=0.358$ to ICC$(2,3)=0.625$ quantified the gain from averaging three partially concordant judgments. The corresponding consistency ICC for the mean of three raters was 0.654, modestly higher than absolute agreement because it does not penalize systematic scale offsets among chemists.

\begin{figure}[htbp]
\centering
\includegraphics[width=0.31\textwidth]{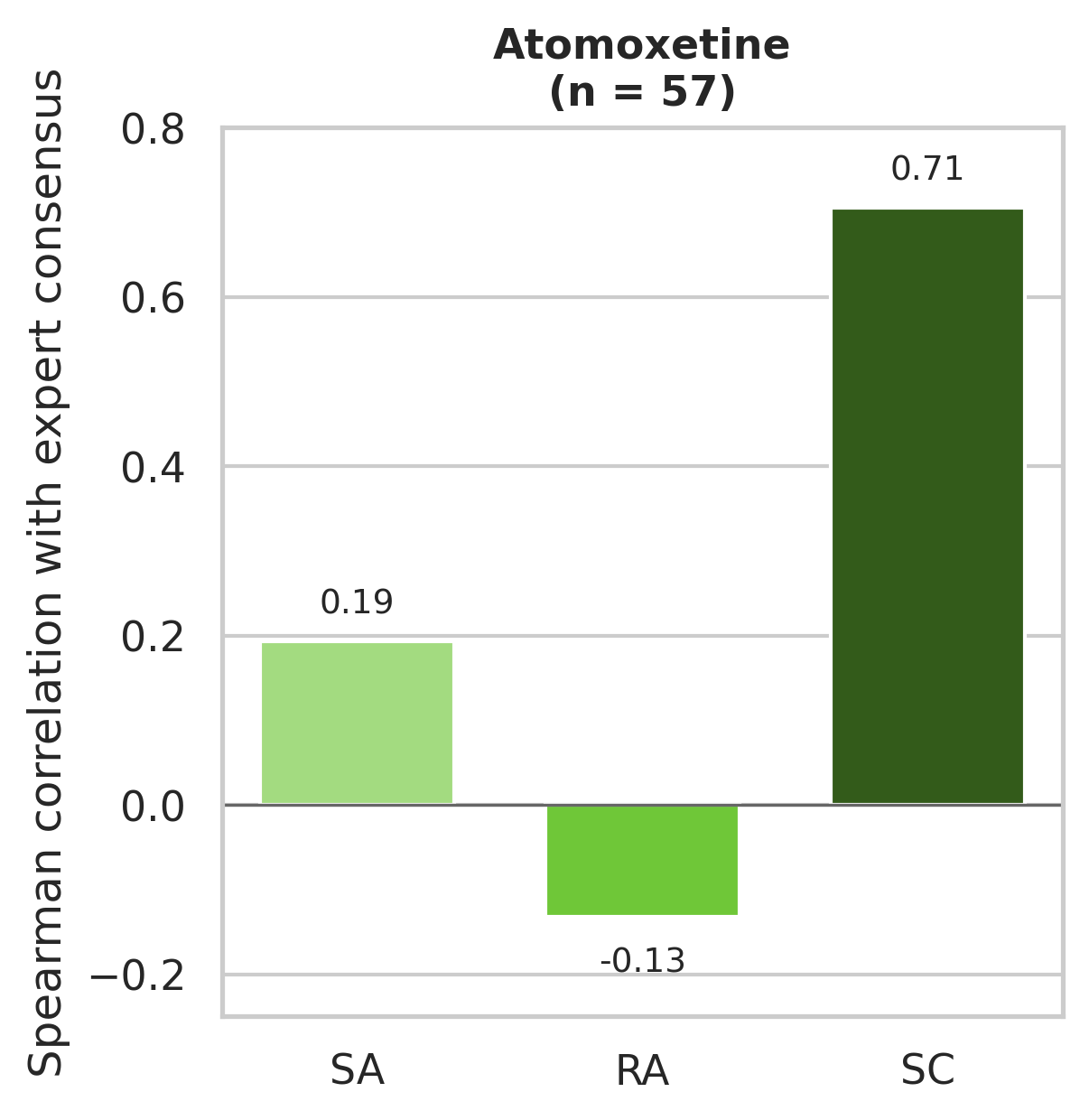}\hfill
\includegraphics[width=0.31\textwidth]{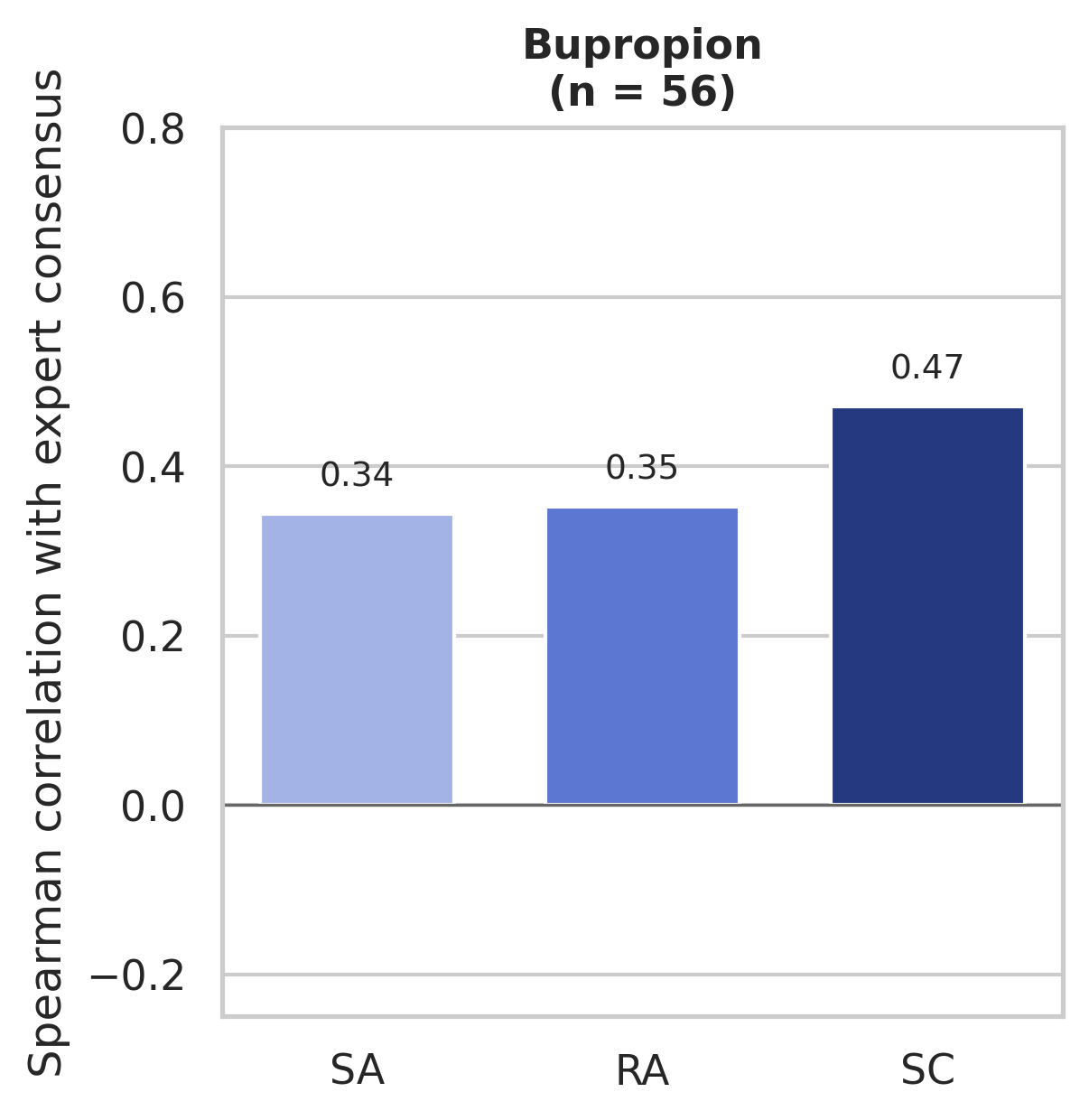}\hfill
\includegraphics[width=0.31\textwidth]{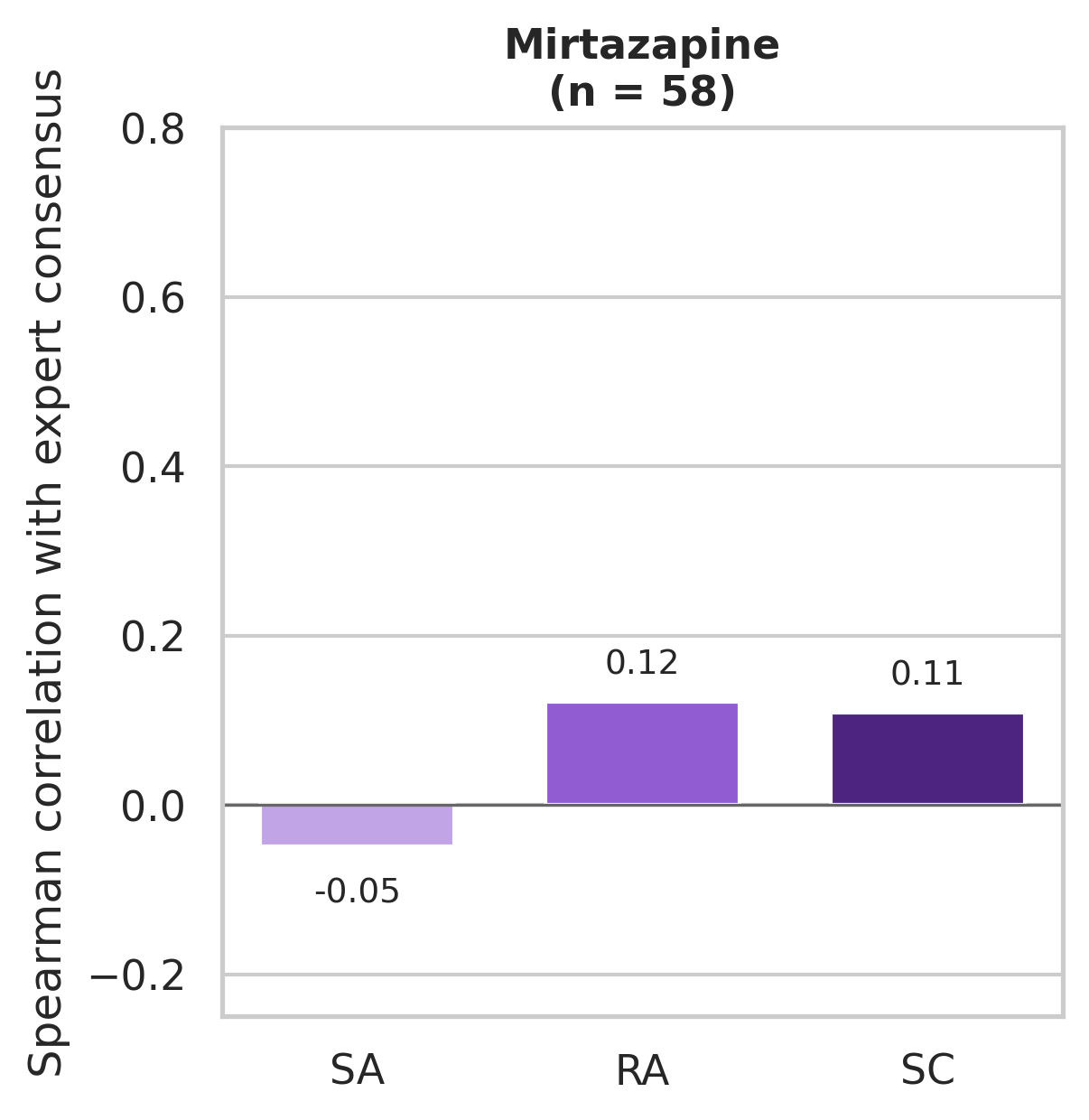}

\vspace{0.8em}
\hfill\includegraphics[width=0.31\textwidth]{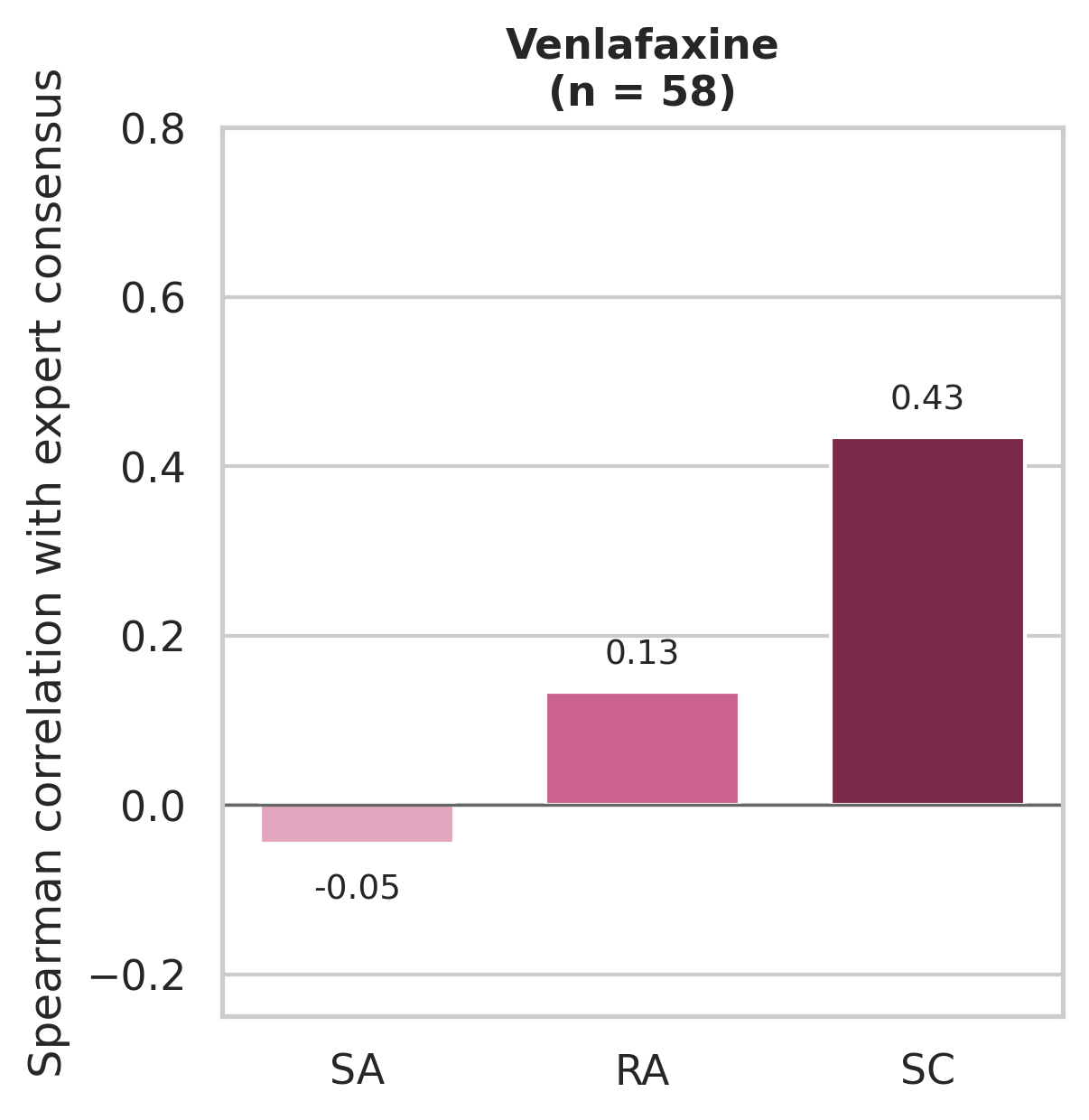}\hspace{0.06\textwidth}
\includegraphics[width=0.31\textwidth]{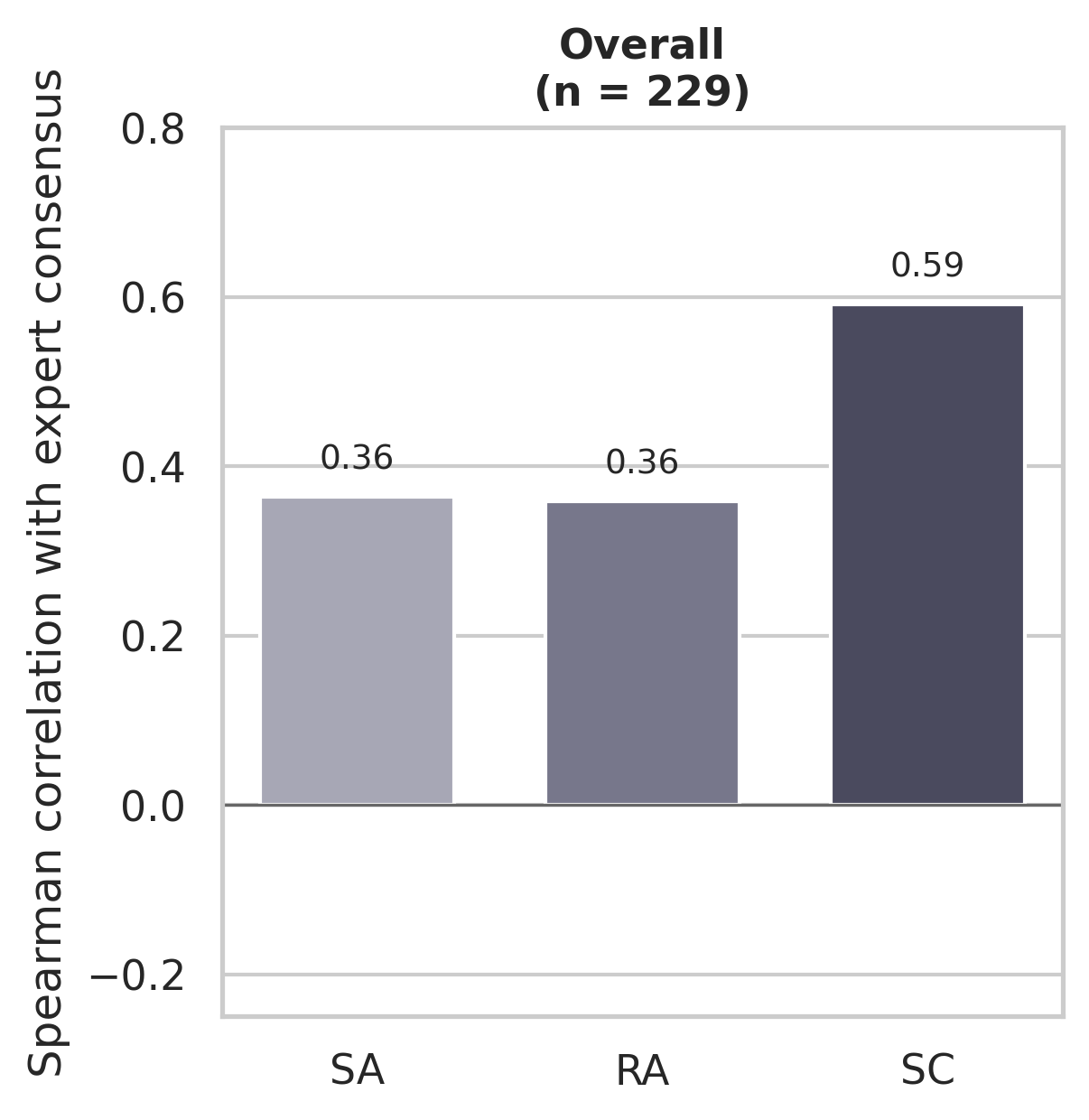}\hfill

\vspace{0.5em}
{\small\color{accent}RAscore is inverted so that higher values indicate greater predicted difficulty.\par}
\caption{Spearman correlation of computational synthesis scores with the blinded expert consensus. RAscore was inverted so that increasing values consistently represented greater predicted difficulty. Correlations were calculated on 229 unique structures.}
\label{fig:expert}
\end{figure}

The computational scores were aligned to a common difficulty direction and standardized over the unique-structure set. Their composite was the arithmetic mean of standardized SA score, inverted RAscore and SCScore. Against the mean expert percentile, the pooled Spearman correlations were 0.364 for SA score, 0.358 for inverted RAscore and 0.591 for SCScore (Fig.~\ref{fig:expert}). The equally standardized composite had a correlation of 0.595. The composite therefore added little rank association beyond SCScore in this sample, although it combined information from three distinct scoring constructions. Standardized Pearson correlations with the consensus were 0.381, 0.307, 0.591 and 0.563 for SA score, inverted RAscore, SCScore and the composite, respectively.

\begin{figure}[!t]
\centering
\includegraphics[width=0.84\textwidth]{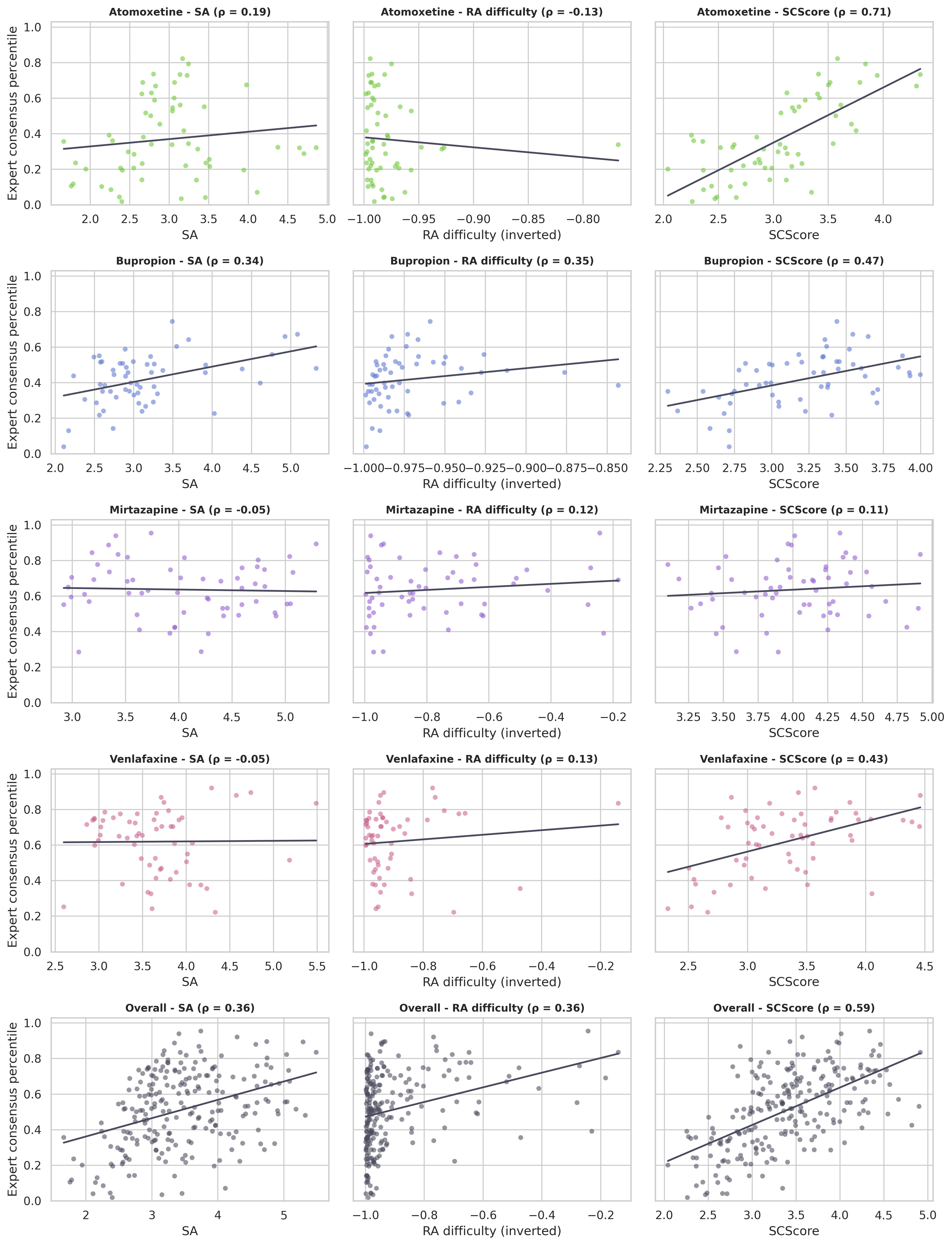}
\caption{Compound-level relationships between SA score, inverted RAscore, SCScore and the blinded expert-consensus percentile. Rows show the four ligand series and the pooled unique-structure set. Lines are descriptive linear fits; panel titles report Spearman correlations.}
\label{fig:expert_scatter}
\end{figure}

Quartile discrimination was expressed as the probability that a randomly selected case from the target quartile received a more extreme algorithmic score than a randomly selected case outside that quartile. For classification of the highest expert-difficulty quartile, AUROC values were 0.631 for SA score, 0.668 for inverted RAscore, 0.728 for SCScore and 0.750 for the composite. For the lowest-difficulty quartile, the corresponding values were 0.711, 0.692, 0.852 and 0.838. Thus, the composite gave the highest discrimination of the hardest quartile, whereas SCScore gave the highest discrimination of the easiest quartile. Compound-level relationships are shown in Fig.~\ref{fig:expert_scatter}.

The pooled associations concealed ligand-specific behavior. SCScore correlated with the consensus at 0.705 for atomoxetine, 0.471 for bupropion, 0.109 for mirtazapine and 0.434 for venlafaxine. The corresponding SA score correlations were 0.193, 0.343, $-0.049$ and $-0.046$. Inverted RAscore correlations were $-0.132$, 0.352, 0.122 and 0.133.

Transfer across ligand series was evaluated by fitting a linear model of expert consensus against standardized SA score, inverted RAscore and SCScore on three ligand series and evaluating it on the fourth. Held-out performance was summarized by Spearman correlation, mean absolute error on the standardized consensus scale and $R^2$ (Table~\ref{tab:loco}). The fitted SCScore coefficient was the largest of the three score coefficients in every fold, ranging from 0.241 to 0.526. Held-out rank correlation remained positive for all four series, but numerical calibration was not preserved for mirtazapine and venlafaxine, for which $R^2$ was below zero.

\begin{table}[htbp]
\centering
\caption{Leave-one-ligand-out calibration of blinded expert difficulty from the three computational scores. MAE is reported on the standardized expert-consensus scale.}
\label{tab:loco}
\small
\begin{tabular}{lrrrrrrr}
\toprule
Held-out ligand & $n$ & Spearman $\rho$ & MAE & $R^2$ & $\beta_{SA}$ & $\beta_{RA}$ & $\beta_{SC}$ \\
\midrule
Atomoxetine & 57 & 0.657 & 0.662 & 0.082 & 0.122 & 0.056 & 0.241 \\
Bupropion & 56 & 0.543 & 0.332 & 0.245 & 0.080 & 0.054 & 0.389 \\
Mirtazapine & 58 & 0.072 & 0.710 & $-1.812$ & 0.150 & 0.210 & 0.526 \\
Venlafaxine & 58 & 0.481 & 0.648 & $-0.679$ & 0.076 & 0.050 & 0.427 \\
\bottomrule
\end{tabular}
\end{table}

The positive held-out rank correlations for atomoxetine, bupropion and venlafaxine indicate partial preservation of ordering, while the negative $R^2$ values show that rank preservation and numerical calibration are separate properties. For mirtazapine, both quantities were weak, consistent with its low score-consensus correlations in the stratified analysis.

\section{Discussion}
The analysis separates four decisions that can otherwise be compressed into a single similarity-search setting. The first is how far to move from a reference ligand. The second is the radial extent of the structural environment encoded by an ECFP representation. The third is whether atom identity or feature-class invariants define the search neighborhood. The fourth is whether computational estimates of synthetic accessibility correspond to an independent human assessment. Each decision altered the composition or interpretation of the retrieved set.

The Tanimoto analysis did not identify a transferable absolute cutoff. Three selected values occurred at the broadest available retrieval boundary, and one occurred at an interior point. Boundary optima do not establish that the lowest observed value is optimal; they show only that further tightening was not favored within the available top-10,000 output. For mirtazapine, the broad near-optimal interval further limits the precision with which a threshold can be stated. These observations favor reporting operating intervals and retained-set sizes together with the numerical cutoff.

The ECFP setting analysis yielded a pooled default but not a universal rule. ECFP8 performed consistently across many priorities, while venlafaxine favored ECFP2. The setting-by-ligand effect was large on the observed utility surface. This result is compatible with the local nature of fingerprint similarity: changing the encoded neighborhood can alter which structural features dominate the ranking, and the effect depends on the reference structure. The cumulative design precludes treating each depth as an independent replicate, so Pareto behavior, rank stability and weight sensitivity provide more direct evidence than nominal model significance.

The ECFP4-FCSFP4 comparison illustrates a related distinction between primary performance and incremental coverage. ECFP4 was more stable under the defined utility, but low compound and scaffold overlap showed that FCSFP4 was not redundant. The venlafaxine search was the clearest case. FCSFP4 contributed 8,134 additional compounds and 2,907 additional scaffolds at the top-10,000 depth, despite not producing a uniform improvement in the synthesis-related metrics. A second-pass search can therefore be useful even when it does not replace the primary fingerprint.

The blinded chemist analysis placed the computational synthesis scores on an external scale. Blinding removes direct anchoring to the algorithms, but it does not make expert judgment an experimental endpoint. Agreement among individual chemists was moderate, and repeat ratings of the same structures varied. Averaging three raters improved reliability, which supports use of a panel consensus rather than a single score. SCScore had the highest pooled correlation with the consensus, although its advantage varied by ligand. The near absence of association for mirtazapine and the negative held-out $R^2$ values for mirtazapine and venlafaxine indicate that pooled correlation alone is insufficient to establish transferability.

Several limitations define the range of the conclusions. The reference set contains four aminergic ligands and should not be taken as representative of all drug-like chemical space. The searches are deterministic ranked outputs rather than randomized samples. WIPO novelty is based on the supplied database annotation and depends on its coverage and scaffold mapping. The utility function assigns equal status to novelty, diversity and synthesis after standardization; weight-sensitivity analysis reduces, but does not remove, that value judgment. The expert benchmark is based on chemical intuition rather than route execution, yield, step count, cost or failed synthesis. Experimental activity was outside the scope of the study.

Taken together, the data support a hierarchical screening procedure. ECFP8 can serve as a starting setting when no ligand-specific calibration exists, and ECFP4 can serve as the primary fingerprint for the present four-ligand panel. The ranked output should then be calibrated to the reference ligand using explicit novelty, diversity and synthesis objectives. FCSFP4 can be added as a deduplicated second pass when nonredundant scaffold coverage is a priority, particularly for venlafaxine-like behavior. Computational synthesis scores can be used for triage, with SCScore providing the closest pooled agreement to the blinded panel, but high-disagreement compounds and poorly transferring ligand series require chemist review. These conclusions concern the construction and prioritization of a screening library. Prospective synthesis and biological testing are required to determine whether the resulting candidates are experimentally accessible and pharmacologically useful.

\section{Methods}
\subsection{Similarity searches and reference ligands}
Similarity searches were initiated from atomoxetine, bupropion, mirtazapine and venlafaxine against the Enamine make-on-demand chemical space using SpaceMACS, which performs maximum common induced substructure searches directly within non-enumerated combinatorial compound spaces \citep{schmidt2022}. Ranked outputs contained up to 10,000 compounds per query. The records used for subsequent analyses included SMILES, rank, Tanimoto coefficient, fingerprint setting, scaffold annotation, WIPO novelty status and three synthetic-accessibility scores.

\subsection{Fingerprint representations and matched comparisons}
The ECFP setting analysis retained the labels ECFP0, ECFP2, ECFP4, ECFP6 and ECFP8 as supplied. To avoid conflating software notation with the underlying radius implementation, these labels are described as settings. ECFP4 and FCSFP4 were compared by ranked retrieval depth rather than raw Tanimoto coefficient, because the similarity scale is representation dependent. Matched depths were 1,000, 2,500, 5,000, 7,500 and 10,000 compounds.

\subsection{Novelty, diversity and synthetic accessibility}
Compound-weighted novelty was the fraction of compounds with a WIPO-novel scaffold among compounds with known novelty status,
\begin{equation}
N=\frac{1}{m}\sum_{i=1}^{m}\mathbb{I}(q_i=\mathrm{novel}),
\end{equation}
where $m$ is the number of compounds with an assigned novelty status, $q_i$ is the scaffold status of compound $i$ and $\mathbb{I}$ is the indicator function. Scaffold diversity was quantified by normalized Shannon entropy \citep{shannon1948}:
\begin{equation}
D=\frac{-\sum_s p_s\log p_s}{\log n},
\end{equation}
where $p_s$ is the proportion of compounds assigned to scaffold $s$ and $n$ is the number of retained compounds. Synthetic accessibility was represented by SA score \citep{ertl2009}, RAscore \citep{thakkar2021} and SCScore \citep{coley2018}. Favorable directions were lower SA score, higher RAscore and lower SCScore. Within each ligand and comparison stratum, the synthesis composite was
\begin{equation}
S=\frac{Z(-SA)+Z(RA)+Z(-SCScore)}{3}.
\end{equation}
Equal-weight utility was
\begin{equation}
U=\frac{Z(N)+Z(D)+S}{3},
\end{equation}
where $N$ is compound-weighted novelty and $D$ is normalized Shannon diversity.

\subsection{Pareto and weight-sensitivity analyses}
A candidate was Pareto-efficient when no other eligible setting was at least as favorable on novelty, diversity and synthesis and strictly better on one or more objectives. For the threshold analysis, candidates were required to retain at least 1,000 compounds. Near-optimal intervals contained Pareto-efficient candidates within 0.15 standardized utility units of the maximum. Weight sensitivity used 5,000 vectors sampled from a uniform Dirichlet distribution,
\begin{equation}
(w_N,w_D,w_S)\sim \operatorname{Dirichlet}(1,1,1), \qquad U_w=w_NN+w_DD+w_SS.
\end{equation}
Winner share was the fraction of sampled weight vectors for which a setting had the highest utility.

\subsection{Expert synthetic-difficulty benchmark}
Three chemists independently scored compounds using chemical intuition and were blinded to SA score, RAscore and SCScore. The three files were joined by fingerprint, query structure, result identity and rank. Within-rater percentile ranks were averaged to produce the primary ordinal consensus, and within-rater standardized values were averaged for calibration analyses. Repeated structures were averaged for algorithm validation. Absolute-agreement intraclass correlations were calculated for single ratings and the mean of three ratings. Kendall's $W$ quantified rank concordance. Algorithm agreement used Spearman correlation,
\begin{equation}
\rho_s(x,e)=\operatorname{cor}\!\left(\operatorname{rank}(x),\operatorname{rank}(e)\right),
\end{equation}
where $x$ is a computational score aligned to increasing predicted difficulty and $e$ is the expert-consensus percentile. Hardest- and easiest-quartile AUROC and leave-one-ligand-out linear calibration were used as complementary assessments.

\subsection{Statistical interpretation}
Ranked top-$N$ sets are nested and are not independent experimental replicates. Regression and paired-test $P$ values are therefore reported as descriptive support. Decisions were based primarily on effect direction and magnitude, Pareto efficiency, rank stability, overlap and sensitivity to alternative objective weights. All algorithm comparisons against chemists used the 229 unique-structure set.

\section*{Data availability}
The processed data supporting this study, including the threshold metrics, matched-depth metrics, overlap summaries and expert-benchmark tables, are available at \url{https://github.com/temisobodu/explore-paper}. Availability of the underlying commercial-library records is subject to the terms of the source database.

\section*{Code availability}
The R scripts used for the three screening questions and score-correlation figures, together with the independently executed Python validation workflow, are available at \url{https://github.com/temisobodu/explore-paper}.

\section*{Author contributions}
T.S. conceived the study and coordinated the analysis. T.S., V.C., R.K. and P.O. contributed to study design, chemical interpretation and manuscript review. The authors reviewed the final manuscript.

\section*{Competing interests}
T.S. is affiliated with Attention Labs. The remaining authors declare no competing interests. This statement should be confirmed by all authors before submission.

\section*{Acknowledgements}
The authors thank the chemists who provided independent synthetic-difficulty assessments. No funding information was supplied for this manuscript.

\bibliographystyle{unsrtnat}
\bibliography{references}
\end{document}